\documentclass{emulateapj}

\newcommand{\mbfv}{\mathbf{v}}

\newcommand{\ddt}{\partial_t}

\begin{document}

\title{Cooling, Gravity and Geometry: Flow-driven Massive Core Formation}

\author{Fabian Heitsch\altaffilmark{1}}
\author{Lee W. Hartmann\altaffilmark{1}}
\author{Adrianne D. Slyz\altaffilmark{2}}
\author{Julien E.G. Devriendt\altaffilmark{3}}
\author{Andreas Burkert\altaffilmark{4}}
\altaffiltext{1}{Dept. of Astronomy, University of Michigan, 500 Church St., 
                 Ann Arbor, MI 48109-1042, U.S.A}
\altaffiltext{2}{Oxford University, Astrophysics, Denys Wilkinson Building, Keble Road,
                 Oxford, OX1 3RH, United Kingdom}
\altaffiltext{3}{Universit\'e Claude Bernard Lyon 1, 
                CRAL, Observatoire de Lyon, 9 Avenue Charles Andr\'{e},
                 69561 St-Genis Laval Cedex, France; CNRS, UMR 5574; ENS Lyon}
\altaffiltext{4}{Universit\"ats-Sternwarte M\"unchen, Scheinerstr. 1, 81679 M\"unchen, Germany}
\lefthead{Heitsch et al.}
\righthead{Flow-driven Core Formation}

\begin{abstract}
We study numerically the formation of molecular clouds in large-scale colliding flows
including self-gravity. The models emphasize the competition between the effects of gravity
on global and local scales in an isolated cloud. Global gravity builds up large-scale
filaments, while local gravity -- triggered by a combination of
strong thermal and dynamical instabilities -- causes cores to form. 
The dynamical instabilities give rise to a local focusing of the
colliding flows, facilitating the rapid formation of massive protostellar
cores of a few $100$~M$_\odot$.
The forming clouds do not reach an equilibrium state, though the motions within
the clouds appear comparable to ``virial''. The self-similar core mass
distributions derived from models with and without self-gravity indicate
that the core mass distribution is set very early on during the cloud
formation process, predominantly by a combination of thermal and dynamical
instabilities rather than by self-gravity.
\end{abstract}

%
%
\section{Rapid Star Formation}\label{s:introduction}

There is increasing evidence that star formation in the solar neighborhood
follows rapidly upon molecular cloud formation 
(\citealp{2001ApJ...562..852H}; \citealp{2007RMxAA..43..123B} and references therein).
This evidence suggests that the density enhancements in which stars form are produced
during the cloud formation phase; thus understanding cloud formation
is essential to understanding star formation.  Moreover, it appears that non-linear
density perturbations need to arise quite early in cloud formation,
as massive, finite molecular clouds are highly susceptible to large-scale
gravitational collapse which could overwhelm small, stellar-mass fragmentation
\citep{2004ApJ...616..288B}.  While several investigations have adopted various
assumed forms of initial and/or driven turbulent motions to produce the necessary
small-scale structure (\citealp{2000ApJS..128..287K}; \citealp{2002ApJ...576..870P};
\citealp{2002MNRAS.332L..65B,2003MNRAS.339..577B}),
it is preferable to have these structures arise naturally.
Thus, a close look at instabilities in cloud
formation which could lead to strong density fluctuations is needed.

\citet{1999ApJ...527..285B} and \citet{2001ApJ...562..852H}
proposed that cloud formation as the result
of pileup of material by large-scale flows is an essential mechanism for explaining
the ``crossing time problem'', i.e. the observation that the typical age spreads in
the stellar populations of many large star-forming regions are often substantially smaller  
than the lateral crossing timescales; in the large-scale flow picture, no information
is transmitted laterally, i.e. perpendicular to the large-scale flow.
This picture works only if star formation follows closely
upon molecular gas formation. Expanding H II regions,
supernova bubbles, and spiral density waves are all obvious candidates for large-scale
supersonic flows which can sweep up material, and there is considerable direct observational
evidence for rapid star formation in these environments (\citealp{2001ApJ...562..852H} and
references therein).  Thus a plausible place to look for stellar core-forming instabilities
is in the post-shock material of the large-scale flows.

Several numerical studies relevant to this problem have now been undertaken
(see the discussions of the literature in \citealp{2006ApJ...648.1052H}
and \citealp{2007ApJ...657..870V}).  These calculations typically assume
converging flows to keep the shocked gas within the
computational volume, but this can easily be extended to describe a more
generic situation by recasting the problem in the rest frame of the shock(s). 
These models overcome the limitations of previous turbulent fragmentation models 
(see review by \citealp{2004RvMP...76..125M})
by avoiding ad hoc assumptions about the source of turbulence and boundary conditions.
Indeed, the converging flow models demonstrate with ease that 
flows provide a natural mechanism for the generation of structure and turbulence in 
clouds (\citealp{2005A&A...433....1A}; \citealp{2005ApJ...633L.113H}; \citealp{2006ApJ...643..245V};
\citealp{2006ApJ...648.1052H}; \citealp{2007A&A...465..445H}; \citealp{2007A&A...465..431H}).

Although this study is motivated by the scenario of molecular clouds being 
transient entities \citep{1999ApJ...527..285B}, forming and dispersing in 
background flows within a few free-fall times, the results presented here 
are not restricted to this scenario. 
Colliding flows can appear even in large scale gravitational instabilities, 
linking our models to the alternative scenario of ``Giant Molecular Clouds'' 
living for substantially longer than only a few free-fall times. The issue 
of cloud lifetimes is currently a matter of debate (e.g. 
\citealp{2007ApJ...654..304K}; \citealp{2007arXiv0707.2252E}, 
\citealp{2007arXiv0707.3514M}), and seems to depend strongly on the 
galactic environment \citep{2001ApJ...562..852H,2006ApJ...648.1052H}. For 
reasons discussed in \S\ref{s:answers}, our models cannot predict cloud life 
times. Thus, the emphasis of this study is on the {\em onset} of star formation.

The purpose of this study is to compare the fragmentation processes in simulated converging 
flows with and without self-gravity in order to show that the rapid
onset of star formation is pretty much unavoidable within the scenario where molecular
clouds form in converging flows.
Recently, \citet{2007ApJ...657..870V} 
presented a study of star formation in clouds formed by colliding flows, emphasizing the long-term
evolution of the system.  Here we focus more on the initial development of the cloud, 
discussing the consequences of cooling, cloud geometry, and gravity 
for the star formation process.

Cooling and gravity both are fragmentation agents, with the difference that gravity can be 
relevant on all scales which surpass the Jeans length, while the isobaric condensation mode of 
the thermal instability \citep{1965ApJ...142..531F} is limited to scales set by the sound speed
and the cooling time, $\lambda_c = c_s\,\tau_c$ 
(e.g. \citealp{2000ApJ...537..270B}; \citealp{2007A&A...465..431H}). Thus, in the early stages
of cloud formation, when only a little mass has  accumulated, the thermal instability is the 
dominant fragmentation agent.

{\em Non-linear} density perturbations collapse at a higher
rate than the global cloud \citep{2004ApJ...616..288B}, 
thus allowing stars to form locally before the whole-sale collapse of the
cloud.  We find that the strong dynamical and thermal instabilities
generate non-linear density perturbations for rapid {\em local} collapse, while still
allowing for the global build-up of the cloud. 

The physics and methods are summarized in \S\ref{s:physmeth}, followed by the
model results in \S\ref{s:results}. Readers solely interested in the discussion
of the results and their consequences should directly proceed to 
\S\ref{s:discussion} and \S\ref{s:answers}.

%
%

\section{Physics and Methods}\label{s:physmeth}

Our study focuses on the effects of global versus local gravity on the one hand, and
on the rapid generation of substructure in the colliding flows on the other.
Since we are interested in global gravitational effects, we cannot use the 
periodic boundary conditions of earlier turbulent fragmentation studies (e.g. 
\citealp{2000ApJ...535..887K}; \citealp{2001ApJ...547..280H};
\citealp{2001ApJ...553..227P}; \citealp{2003ApJ...592..203G}; \citealp{2005ApJ...618..344V}).
Neither will we generate or drive turbulence by imposing a randomly chosen velocity or density field, 
but instead, we will rely on turbulence generated by the dynamical instabilities arising
from the collisions of the flows. 


\subsection{The Models}\label{ss:models}
We ran four models, whose parameters are listed in Table~\ref{t:modparam}.
All models are run on a fixed grid with the instreaming gas flowing along the 
$x$-direction, entering the domain at the $(y,z)$-planes. To trigger
the fragmentation of the (otherwise plane-parallel) interaction region,
we perturb the collision interface. We chose the
perturbations of the collision interface from a random distribution of 
amplitudes in Fourier space with a top hat distribution restricted
between wave numbers $k=1..4$. 

Model Gs (for ``gravity in shell'') can be interpreted as
colliding continuous gas streams in spiral shocks (e.g. \citealp{1990imeg.conf..298T}
for observational evidence, and \citealp{2007MNRAS.376.1747D} for numerical models),
or as a close-up view of two expanding and colliding super-shells in the LMC.
The collision interface is plane-parallel, except for the imposed perturbations. 
Material is free to leave the box in the lateral (i.e. perpendicular to 
the inflow) directions.

In models Gf1 and Gf2 (for ``gravity in finite cloud''), we restrict the 
inflow to a cylinder of elliptical cross section with an ellipticity of $3.3$
and a major axis of $80$\% of the (transverse) box size, 
mimicking two colliding gas streams in a more general geometry. Again, the 
collision interface is perturbed. The motivation here is to generate one
finite cloud in order to study global gravitational effects.

Finally, model Hf1 is a non-gravitating version of Gf1, to compare the role
of gravity versus that of the thermal instability for the fragmentation of the 
gas streams.

The inflow density in all models is $n_0=3$~cm$^{-3}$ at a temperature of
$T_0=1800$~K and an inflow velocity of $7.9$~km~s$^{-1}$, corresponding to
a Mach number of ${\cal M} = 1.5$. The flows are initially in thermal
equilibrium. The models start at time $t=0$ with the collision of the two
flows. For models with spatially constrained inflows (Hf1, Gf1, Gf2), the 
fluid is at rest everywhere except in the colliding cylinders.

\begin{deluxetable}{c|ccccc}
  \tablewidth{0pt}
  \tablecaption{Model Parameters\label{t:modparam}}
  \tablehead{\colhead{Name}&\colhead{$n_xn_yn_z$}
             &\colhead{$L_xL_yL_z$ [pc]}
             &\colhead{gravity}
             &\colhead{$t_{end}$ [Myr]}
             &\colhead{$\eta$ [pc]}}
  \startdata
  Hf1 & $256\times 512^2$ & $22\times 44^2$&no  & 14.5 & $2.2$ \\
  Gf1 & $256\times 512^2$ & $22\times 44^2$&yes & 14.5 & $2.2$ \\
  Gf2 & $256\times 512^2$ & $22\times 44^2$&yes & 14.5 & $4.4$ \\
  Gs  & $256^3$           & $44^3$&yes & 16.0 & $2.2$
  \enddata
  \tablecomments{1st column: Model name. 2nd column: resolution. 3rd column:
                 physical grid size. 4th column: gravity. 5th column: end time of run.
                 6th column: amplitude of interface displacement.}
\end{deluxetable}

The finite cloud models Gf1, Gf2 and Hf1 have
a grid cell size of $\Delta L=8.6\times 10^{-2}$~pc, while model Gs has one of
$\Delta L=1.7\times 10^{-1}$~pc (Tab.~\ref{t:modparam}). We note that this
does {\em not} constitute the physical resolution power of the simulation.
At minimum, the stencils (i.e. the support points) used for the higher-order 
reconstruction of the fluid states at the cell wall will render the cells within one stencil 
not independent. In other words, conclusions should not be drawn from structures of $4$ 
cells or less of linear size. Thus, we use only cores with $64$ or more cells for analysis.

\subsection{Boundary Conditions}\label{ss:boundcond}
The $x$-boundaries are partly or entirely defined as inflow-boundaries,
 depending on the model.
Indeed, the inflow is either defined over the whole
$(y,z)$-plane (model Gs), or within an elliptical surface (models Hf1, Gf1, Gf2:
see \S\ref{ss:models} for details).
The $y$ and $z$ boundaries, -- as well as the part of the $x$-boundaries
that is not occupied by the inflow in models Hf1, Gf1 and Gf2 --, are open, 
meaning material is free to leave
the simulation domain through these boundaries.
This is bound to cause trouble once material tries
to ``come back'' during the later stages of the simulation (note that
this material is not actually coming back, but that it is the result of
the extrapolation of the last active cells properties, i.e. material
with the properties of the last active cell layer within the domain will try
to enter the domain).
This inevitably will happen once global gravity dominates over the
over-pressurized material shooting out of the flow-collision region.
However, the ``re-entering'' material does not reach the central 
cloud region within the simulation time, and in any case contributes only
a negligible amount to the total mass within the box.

The situation becomes more critical once material is leaving the simulation domain
in the $x$-direction, i.e. once it is moving against the inflow. Since the bounding
shocks (and the cooling) will set the density and the temperature of the post-shock
gas, the physical state of the gas will be undefined once the bounding shocks move
off the grid. This will render the ``returning'' material essentially in a 
hydrodynamically inconsistent state. Thus, once material
encounters the $x$-boundaries, the simulation needs to be stopped.

In model Gs, the inflow velocities are slightly reduced at the edges of the domain,
mimicking the velocity profile of an expanding shell of material driven by two 
sources at a distance of approximately $100$ pc to the left and to the right of
the mid plane of the simulation domain. This reduces the amount of material collected
at the edges of the domain, and thus limits the edge effects due to gravity at
later stages of the simulation.

\subsection{Hydrodynamics and Atomic Line Coolants}\label{ss:hydrocool}
As in our previous studies of colliding flows, we used the 
higher-order gas-kinetic grid method Proteus
(\citealp{1993JCoPh.109...53P}; \citealp{1999A&AS..139..199S}; 
\citealp{2004ApJ...603..165H}; \citealp{2005MNRAS.356..737S};
\citealp{2006ApJ...648.1052H}, \citeyear{2007ApJ...665..445H}), allowing
full control of viscosity and heat conduction.
The code evolves the
Navier-Stokes equations in their conservative form to second order in time and
space. The hydrodynamical quantities are updated in time unsplit form.

The heating and cooling rates are restricted to optically thin
atomic lines following \citet{1995ApJ...443..152W}.
Dust extinction becomes important above
column densities of $N(\mbox{HI})\approx 1.2\times 10^{21}$cm$^{-2}$, which are
only reached in the densest regions modeled. Thus, we use the unattenuated
UV radiation field for grain heating \citep{1995ApJ...443..152W},
expecting substantial uncertainties in cooling rates only for the densest regions.
The ionization degree is derived from a balance between ionization by cosmic rays and
recombination, assuming that Ly $\alpha$ photons are directly reabsorbed.
Numerically, heating and cooling is implemented iteratively as a source
term for the internal energy $e$ of the form
\begin{equation}
  \ddt e = n\Gamma(T) - n^2\Lambda(T)\,[\mbox{erg}\mbox{ cm}^{-3}\mbox{ s}^{-1}].
  \label{e:cooling}
\end{equation}
Here, $\Gamma$ is the heating contribution (mainly photo-electric heating from grains),
$n\Lambda$ the cooling contribution (mainly due to the CII HFS line at $158\mu$m).
Since the cooling and heating prescription has to be added outside the
flux computations, it lowers the time order of the scheme.
To speed up the calculations, equation~(\ref{e:cooling}) is tabulated on a $2048^2$ grid
in density and temperature. For each cell and iteration, the actual energy change
is then bi-linearly interpolated from this grid.

The cores forming due to gravitational collapse reach densities of a few 
$10^5$~cm$^{-3}$, far beyond the applicable range of our cooling
curve. Strictly speaking, we should therefore 
extend the cooling curve to include molecular lines at high densities.
However, our cooling curve reaches an equilibrium temperature of approx $12$K for 
$n>10^3$~cm$^{-3}$, close enough to a realistic temperature for molecular cores. 
Since we cannot resolve the core structure anyway, we chose to stick 
to this simplified treatment.

The cooling curve is limited to densities of $n\leq 10^5$~cm$^{-3}$, to prevent a
``catastrophic'' collapse which would be generated by a sub-isothermal effective equation of state.
The sudden reversal to an adiabatic equation of state helps limit the densities
and prevents numerical artifacts caused by single, very high density cells. 
However, this stiffening of the equation of state -- if introduced at too low densities -- 
could stabilize the cores, prevent their fragmentation and render them
more prone to dispersion. We experimented
with the density threshold $n_{max}$ and found that a value 
of $n_{max}=10^5$~cm$^{-3}$ prevents the run-away collapse while allowing
the cores to remain small (and dense) enough to stay gravitationally 
bound once they have formed. See \S\ref{ss:resolution} for a discussion of the 
resolution limits.

\subsection{Gravity}
Self-gravity is implemented as an external source term in time-unsplit form.
The Poisson-equation is solved via a non-periodic Fourier solver, using
the (MPI-parallelized) {\tt fftw} (Fastest Fourier Transform in the West) 
libraries. We tested this against direct summation 
to assure that the Poisson equation is solved accurately. 
The non-periodic solver needs twice the grid size for 
padding the Fourier transforms. This limits the resolution
of our simulations to effectively $512^3$ cells.

\subsection{Core Identification}\label{ss:coreident}

We use two methods to identify cores in the model data.
To find gravitationally bound  objects,
we employ the CLUMPFIND algorithm \citep{1994ApJ...428..693W}
in a modified version \citep{2000ApJ...535..887K}. We then test 
whether the structures identified by CLUMPFIND are gravitationally bound or collapsing,
by checking their Jeans mass, the ratio of thermal and (internal) kinetic
energy over gravitational energy, and the velocity divergence.
If all three tests are passed, a structure is accepted as a core.
We track individual cores by identifying the closest ``neighbor'' to a given
core in the next timestep (where ``timestep'' does not mean the CFL-timestep, 
but the time between writing data sets). The simulations presented here form
sufficiently few cores for this simple method to be accurate.

The second method is a simple clipping algorithm, motivated by
the fact that due to the thermal instability, dense coherent
regions are generally well defined in our models. The method
selects the maximum density and builds a tree structure around
the central cell, thus connecting all cells above the given 
density threshold of $n_{th}>50$~cm$^{-3}$. Once this threshold is reached,
the process restarts with the next-lower density
peak not included in the previous structure. The resulting cores
are accepted independently of whether they are gravitationally bound or collapsing.

%
%
\section{Model Results}\label{s:results}

The general signature of fast local fragmentation in colliding flows is most easily
recognized in the morphologies of the clouds (\S\ref{ss:morph}). A more quantitative
measure can be gleaned from the core mass evolution and the energy distribution 
(\S\ref{ss:colhist}). Dynamical signatures are discussed in \S\ref{ss:gasdyn},
and \S\ref{ss:resolution} sounds two cautionary notes regarding the numerical resolution.

\subsection{Morphologies}\label{ss:morph}

We begin by comparing the morphologies of the clouds forming in 
colliding flows (models Hf1, Gf1 and Gf2, \S\ref{sss:global}).
A new formation mechanism for massive cores is discussed in \S\ref{sss:dynafocus} .

\subsubsection{Global Collapse and Filament Formation}\label{sss:global}
\begin{figure*}
  \includegraphics[width=\textwidth]{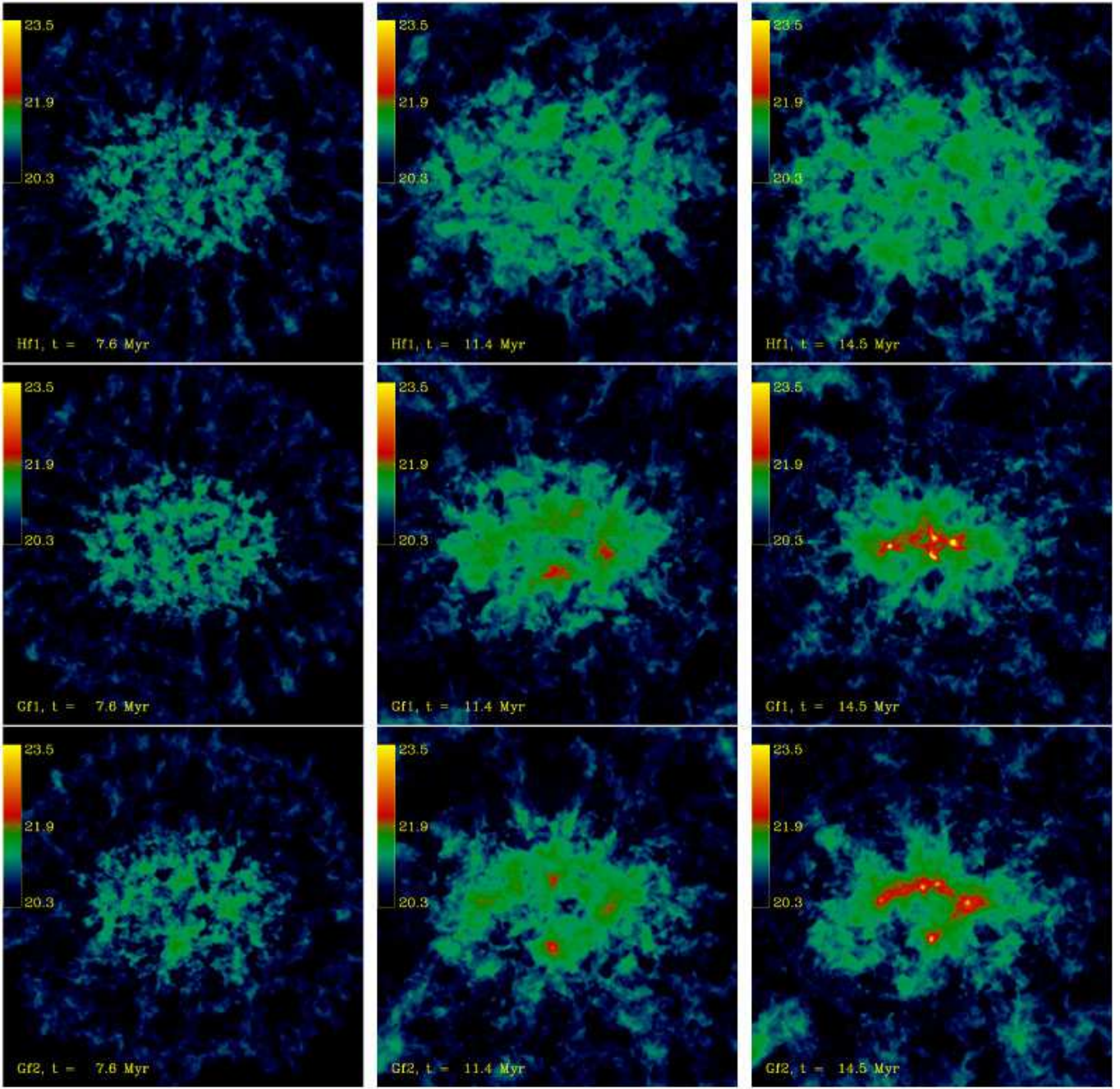}
   \caption{\label{f:morph-Gf12l}Time sequence of logarithmic column density maps
            for models Hf1 (top), Gf1 (center) and Gf2 (bottom), seen along the inflow direction.
            At $t=7.6$~Myr (left column) the full domain (44~pc) is shown, while we restrict the field
            of view to the central 3/4 (33~pc) of the domain at later times, to highlight the small-scale
            dense structures forming.}
\end{figure*}

The top row of Figure~\ref{f:morph-Gf12l} shows three
time instances of model Hf1, seen along the inflow direction. The colliding
flows cause a big ``splash'', the effects of which are still noticeable $7.6$~Myr
after the initial flow collision (left column), however, rapid cooling leads to strong density
enhancements in the interaction zone, and in combination with the dynamical
instabilities triggered by the perturbed interface, to strong fragmentation.
Note the radial filaments and the outermost ``bounding ring'' at $7.6$~Myr. The radial
filaments are similar to those seen in the conceptually equivalent models by
\citet{2007ApJ...657..870V}: the compressed material is escaping the interaction
region by the way of least resistance, i.e. laterally to the inflow. Once 
an ``escape channel'' has been formed due to small fluctuations in the external
pressure, the resulting pressure deficit will ensure that the channel will continue
to be used by subsequent material. The ring-like structure is just
the shock wave from the initial splash caused by the colliding flows. 

With increasing time, more and more mass is collected in the interaction region,
rapidly assembling the cloud. Note that for the sake of greater detail 
the center and right-hand column of
Figure~\ref{f:morph-Gf12l} show only 3/4 (in linear extent) of the simulation domain
corresponding to a box length of 33~pc. The non-gravitating model Hf1 also continues  
to collect mass as time proceeds, however the column densities reached do not exceed a 
few $10^{21}$~cm$^{-2}$.

\begin{figure}
  \begin{center}
  \includegraphics[width=\columnwidth]{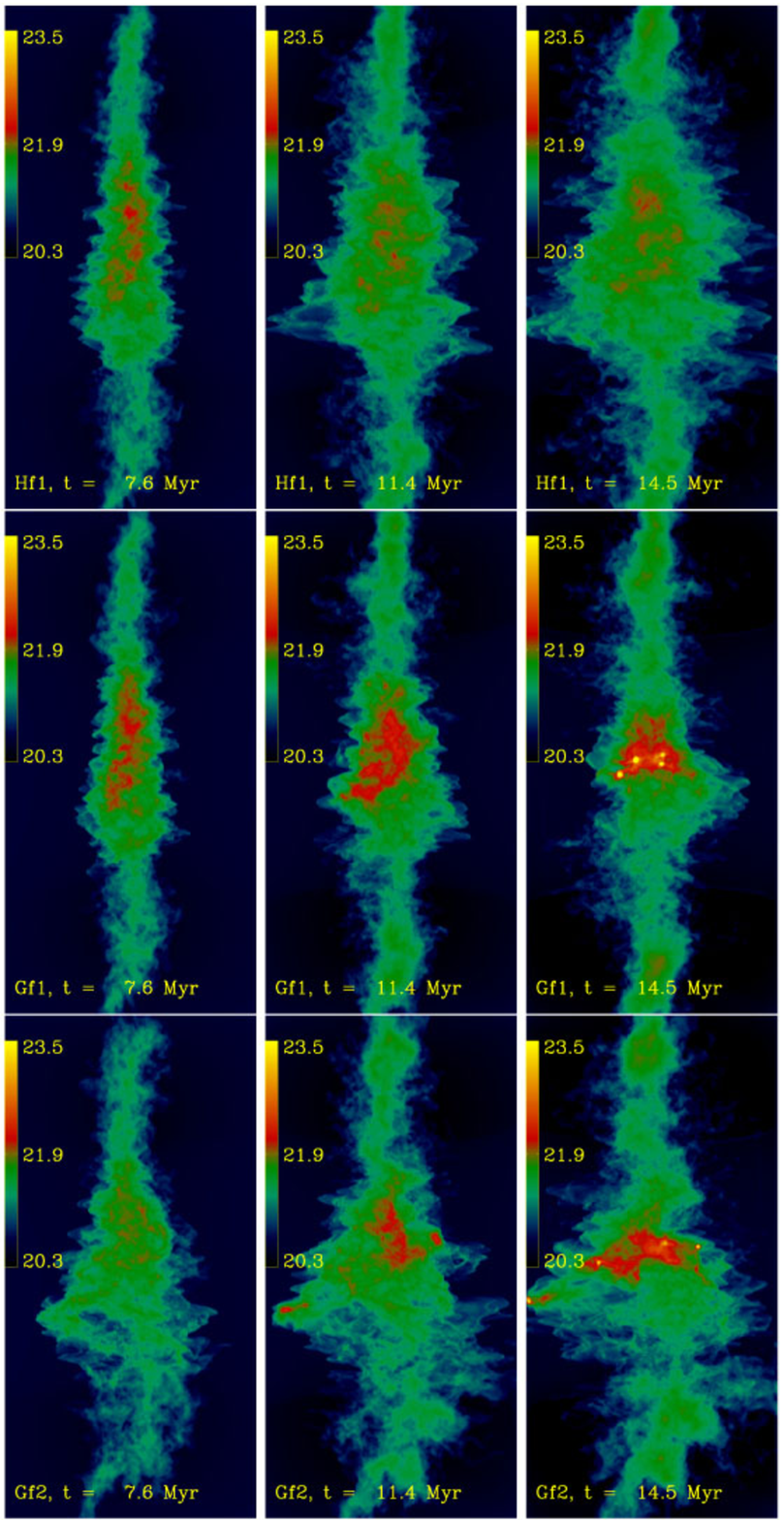}
  \end{center}
   \caption{\label{f:morph-Gf12p}Time sequence of logarithmic column density maps
            for models Hf1 (top), Gf1 (center) and Gf2 (bottom), seen perpendicular to the 
            inflow direction. The full computational domain is shown, measuring $22\times 44$~pc.}
\end{figure}

This changes with the introduction of gravity (center and lower row of Fig.~\ref{f:morph-Gf12l}).
At $7.6$~Myr, the structures are still pretty much indiscernible, while at $11.4$~Myr, the first
regions of high column density have formed. 
While the cores forming in the cloud result from local collapse (see below), the filaments
at later stages ($14.5$~Myr) are a consequence of the global collapse of the whole cloud. This can be
more easily seen in the right-hand column of Figure~\ref{f:morph-Gf12p}. The same models
at the same times as in Figure~\ref{f:morph-Gf12l} are shown, but now seen perpendicularly
to the inflow. Clearly, at late times, the initially elongated ``red'' structure (the dense
part of the cloud) crumples under its own weight. Because of the (generalized) non-circular inflow
cross section, this leads to the formation of a filament. This mechanism for filament formation
offers a substantial reservoir of mass for further star formation. The subsequent fragmentation
of the filaments is possibly enhanced by the thermal instability 
(see e.g. \citealp{2001Ap&SS.276.1097T}).

The larger amplitude of the collision interface perturbation in model Gf2 (see Tab~\ref{t:modparam})
leads to a stronger initial fragmentation, mirrored in a more distributed core formation
at later stages.

\begin{figure*}
  \begin{center}
  \includegraphics[width=\textwidth]{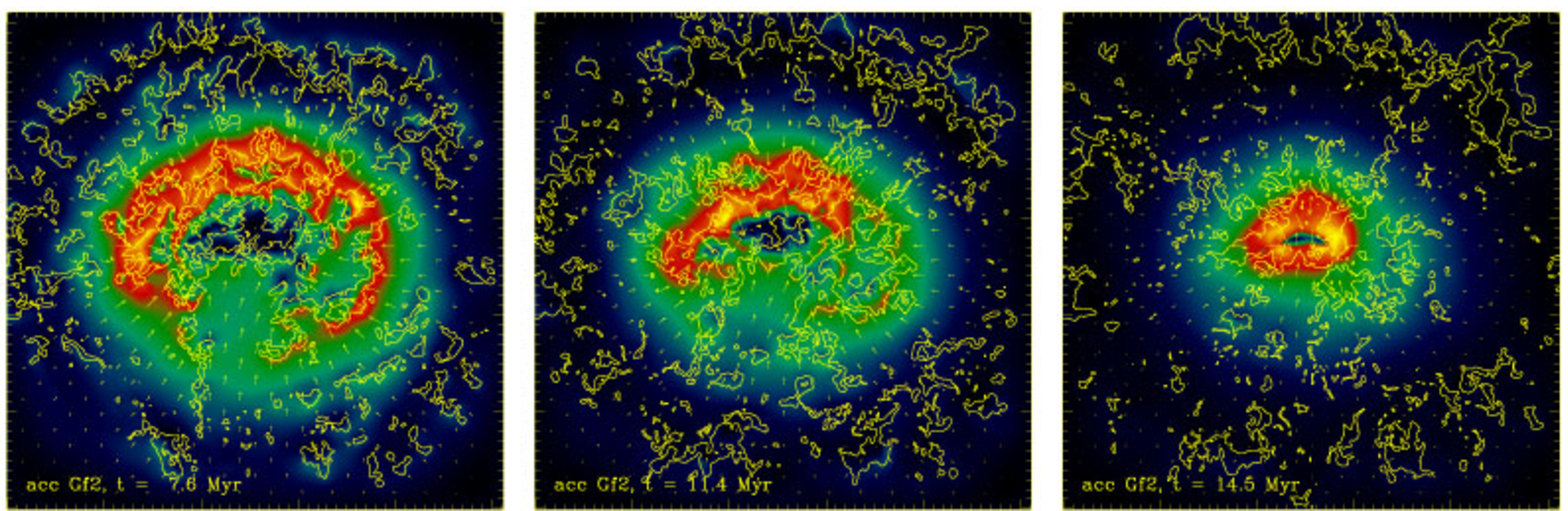}
  \end{center}
   \caption{\label{f:accel3d}Time sequence of accelerations for model Gf2, projected in plane perpendicular to the
            inflow. Colors denote $|\nabla\Phi|$, the contours represent density, and the arrows
            indicate $\nabla\Phi$. This should be compared to the bottom row of Figure~\ref{f:morph-Gf12l}.}
\end{figure*}

Note that the high-density regions (in Fig.~\ref{f:morph-Gf12l}) do not necessarily
form at the center of the cloud. On the contrary, there is a tendency for material to collect
away from the center, forming filaments (model Gf2 at $14.5$~Myr), or at least extended dense cores.
This is a mild version of the edge effect in collapsing finite sheets, as discussed by
\citet{2004ApJ...616..288B} and applied to a model of the Orion star forming region
by \citet{2007ApJ...654..988H}. Figure~\ref{f:accel3d} is more specific about the actual mechanism:
It shows a map of the (projected) gravitational accelerations $|\nabla\Phi|$
in the midplane perpendicular to the inflow (to be
compared to the bottom row of Fig.~\ref{f:morph-Gf12l}). Contours denote the column density, and
the actual potential gradient $-\nabla\Phi$ is indicated by the arrows. The color table is identical
to Figure~\ref{f:morph-Gf12l}, i.e. yellow/red indicates strong accelerations. The ring-like structure 
of strong accelerations towards the center is obvious. Note that the accelerations extend towards larger
radii than their accompanying high-density structures: material is accelerated at the edges and is
being piled up further down the (radial) flow. 

The simulations of \citet{2007ApJ...657..870V} show a similar
effect as the models discussed here, although much stronger. This quantitative difference
might be a consequence of the choice of initial perturbations in the two simulation sets. 
While \citet{2007ApJ...657..870V} put small-scale perturbations in their inflow velocities,
we perturb the collision interface, which in turn leads to a stronger excitation of dynamical
instabilities. The predominant instability
arising in our setup is the non-linear thin shell instability (NTSI, \citealp{1994ApJ...428..186V}),
in combination with shear flow instabilities and the thermal instability \citep{1965ApJ...142..531F},
preventing the formation of a more or less uniform slab. 


\subsubsection{Dynamical Focusing and Massive Core Formation}\label{sss:dynafocus}

The NTSI is driven by the ram pressure imbalance at the troughs of 
the rippled interaction surface: the concave side will have an excess of ram pressure
because the flows are focused into the troughs, while the convex side will 
experience a deficit of ram pressure because the incoming gas is deflected.
Thus, the NTSI provides a very efficient mechanism to collect gas locally.
\citet{2003NewA....8..295H} demonstrated this effect with
the help of self-gravitating two-dimensional models.

\begin{figure}
  \includegraphics[width=\columnwidth]{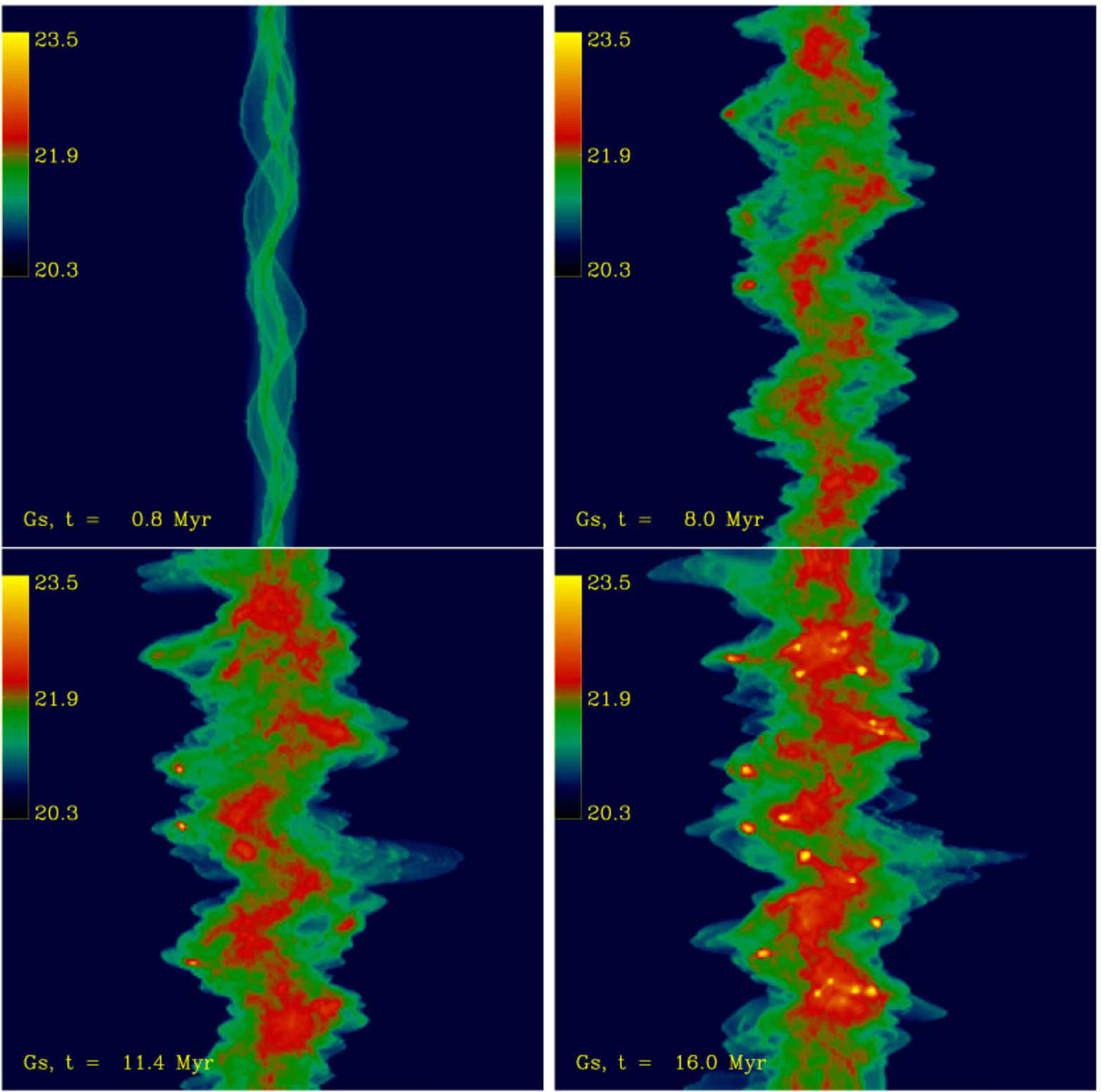}
  \caption{\label{f:morph-Gs}Time sequence of logarithmic column density maps
           for model Gs, seen perpendicular to the inflow direction. The domain measures
           44~pc in linear extent.}
\end{figure}

This focusing effect still holds in three dimensions, as can be seen from 
the time sequence of logarithmic column density maps of two colliding flows
under the effect of self-gravity, as shown in Figure~\ref{f:morph-Gs}. 
The leftmost panel at $t=0.8$~Myr shows structures still pretty close to the
initial conditions. The cooling has not yet led to perceptible fragmentation,
and turbulence has not yet developed. At $8.0$~Myr, the NTSI is already in 
full swing, and on the left edge of the cloud, roughly in the mid-plane, the first
core starts to form. This core is located at one of the troughs amplified
by the NTSI, so that it had ample opportunity to collect instreaming material.
This seems to continue all the way up to $16$~Myr, at which point the cloud
has grown globally unstable (see \S\ref{ss:colhist}), indicated by the 
frenzy of core formation all over the cloud. We will present a more
detailed discussion of the dynamical focusing in a subsequent paper,
including models at higher resolution.

\begin{figure}
  \includegraphics[width=\columnwidth]{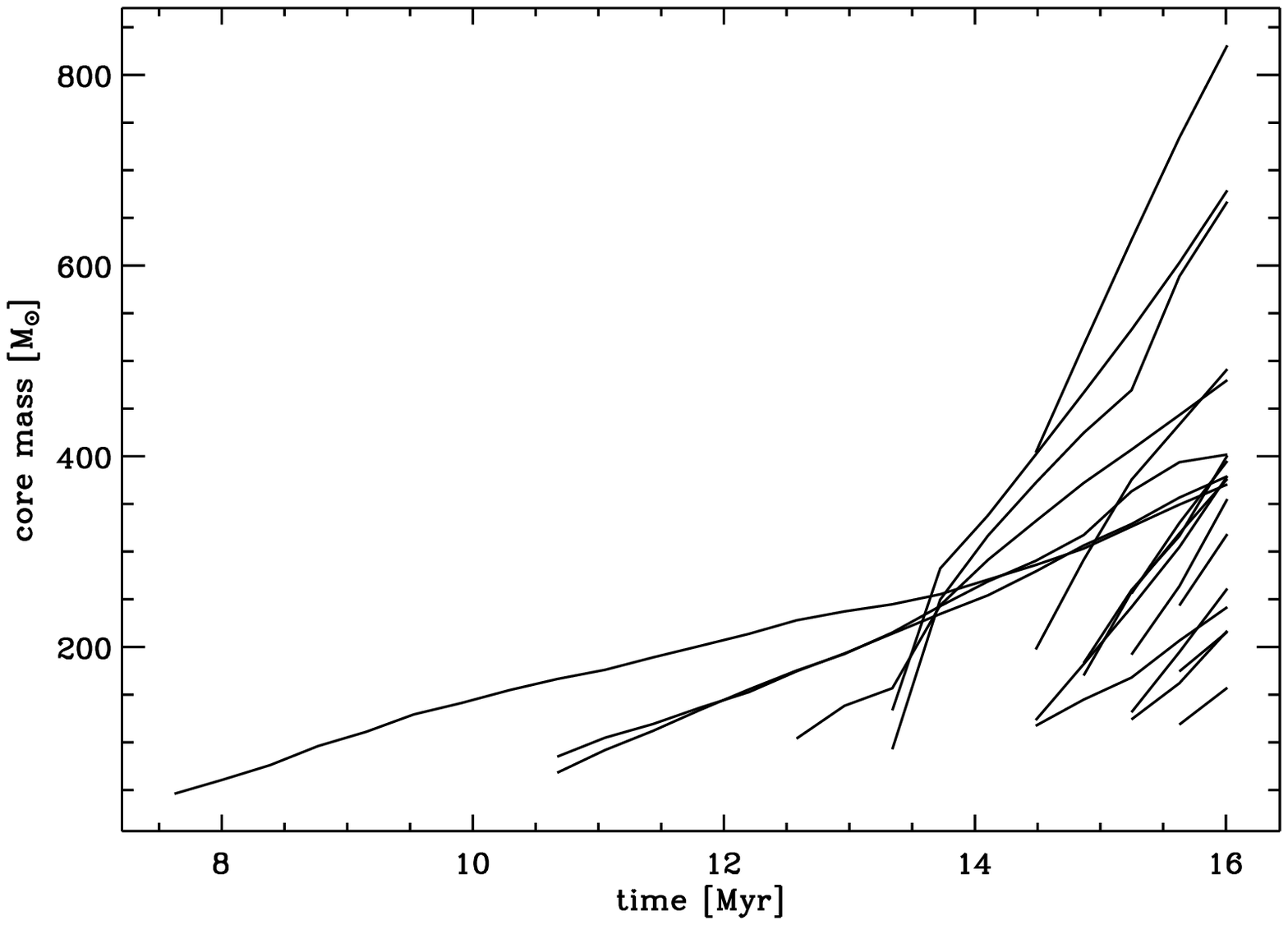}
  \caption{\label{f:masstime-Gs}Core mass evolution for model Gs. The global
           collapse of the cloud is paralleled by a frenzy of core formation
           at late times.}
\end{figure}

Isolating the gravitationally collapsing cores with CLUMPFIND (see \S\ref{ss:coreident}) 
and tracking the core masses with time yields Figure~\ref{f:masstime-Gs}. 
The dynamical focusing seems to be a very efficient
mechanism to collect substantial mass in a small volume over a short time: 9~Myr
after initial flow contact, the first $100$~M$_\odot$-core has formed. At later
times, the mass accretion rates get steeper: there is more mass available, and 
(concurrently) the potential well deepens, so that more massive cores can form
within shorter times. The first core does not participate in the steepened
accretion history because it sits at the edge of the potential well so 
that it does not benefit from the higher densities. 

Obviously, stellar feedback will influence the cloud dynamics and star formation
efficiency at the late stages of the cloud evolution. 

\subsection{Collapse History}\label{ss:colhist}

\subsubsection{Energy Equipartition}\label{ss:equip}

\begin{figure*}
  \includegraphics[width=\textwidth]{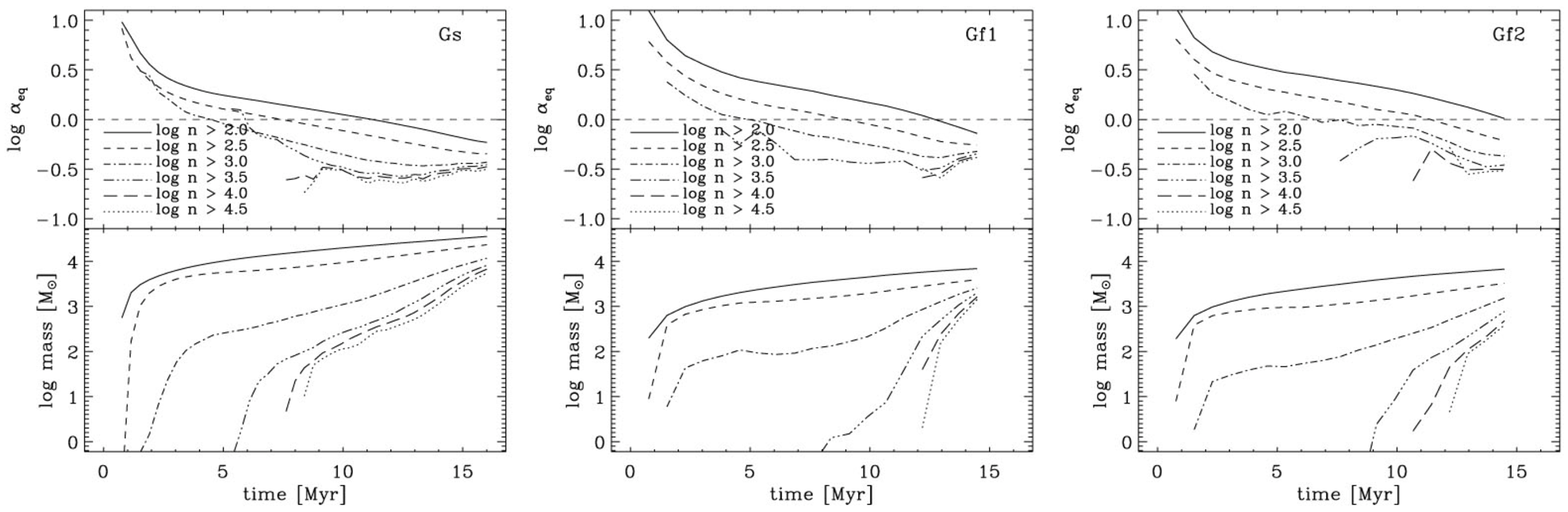}
  \caption{\label{f:virpar}The equipartition parameter $\alpha_{eq}$ against time for
           the three gravity models Gs, Gf1 and Gf2 as indicated in the panels. The top half of each panel
           gives $\alpha_{eq}(n>n_{th})$, the bottom half shows the total mass $M(n>n_{th})$.
           The line styles stand for the threshold densities $n_{th}$ as indicated in the top panels.}
\end{figure*}

A global measure for the cloud evolution under the effect of gravity is the 
equipartition parameter 
\begin{equation}
  \alpha_{eq} \equiv \frac{\int(\rho\mbfv^2+3P)dV}{\int\rho\Phi dV},\label{e:alphaeq}
\end{equation}
i.e. the ratio of the total kinetic and thermal energy over half the total potential 
energy\footnote{We decided to follow \citet{2006MNRAS.372..443B} and replace the term ``virial
parameter'', since this usually has the connotation of ``virial equilibrium'', an assumption,
which generally holds only for an ensemble of molecular clouds, or for a time-average
of one cloud over many dynamical times \citep{1999osps.conf...29M}. Since we are concerned
here with a single molecular cloud on short timescales, the notion of virial equilibrium is
inapplicable -- while that of energy equipartition may still hold \citep{2006MNRAS.372..443B}.}.
We neglect any surface terms in the determination of $\alpha_{eq}$,  which only allows us to obtain a
rough approximation of the ``true'' energetic state of the cloud. Nevertheless, as a closer study 
of Figure~\ref{f:virpar} demonstrates, the time evolution of $\alpha_{eq}$ mirrors the
cloud morphologies (Figs.~\ref{f:morph-Gs} and \ref{f:morph-Gf12l}).
Figure~\ref{f:virpar} shows $\alpha_{eq}$ for all the gas above a given threshold density. 
A low threshold density means that most of the mass and most of the volume enter 
the calculation of $\alpha_{eq}$.
Increasing the density threshold emphasizes more and more the dense cores that form later. 
The lowest density threshold is set to $n_{th}=10^{2}$~cm$^{-3}$, since we are interested
in the equipartition parameter of the isolated cloud, and not of the total simulation volume
including the (highly energetic) inflows. The following items are noteworthy:

(1) The dense regions tend to be gravitationally
bound ($\alpha_{eq} < 1$), while the global cloud behavior (illustrated by the lowest-density
curves) approaches $\alpha_{eq} \sim 1$ by the end of the simulations.  Thus, in terms of
observables, the cloud exhibits an $\alpha_{eq}$ consistent with ``virial equilibrium'' to
within a factor of two, well within observational uncertainties in terms of mass estimates,
velocity dispersions derived only from line-of-sight motions, and the elimination of
surface terms (\citealp{1999ApJ...515..286B}; \citealp{2006MNRAS.372..443B}).

(2) The solid lines indicating the lowest density threshold of $n_{th}=10^2$~cm$^{-3}$
reach $\alpha_{eq}=1$ between $10$ and $13$~Myr. A lower density threshold can be interpreted
as tracing a larger volume, thus the evolution of $\alpha_{eq}$ for low density thresholds
indicates that global collapse lags behind the local collapse -- isolated
dense cores form before the cloud can collapse globally. However, the cloud
{\em does} show the onset of global collapse. 

(3) At high density thresholds, $\alpha_{eq}<1$ for all times (at which high
densities are available): the massive cores (see the corresponding mass history)
are fully gravitationally unstable. Note that $\alpha_{eq}$ does not drop further,
although the mass increases: the cut-off of the cooling curve leads to a stabilization 
of the cores and prevents catastrophic collapse. 

(4) The clouds do not go through an ``equilibrium'' stage, but start to collapse
locally during their formation: stars can form locally without a global collapse
of the cloud.

\begin{figure*}
  \includegraphics[width=\textwidth]{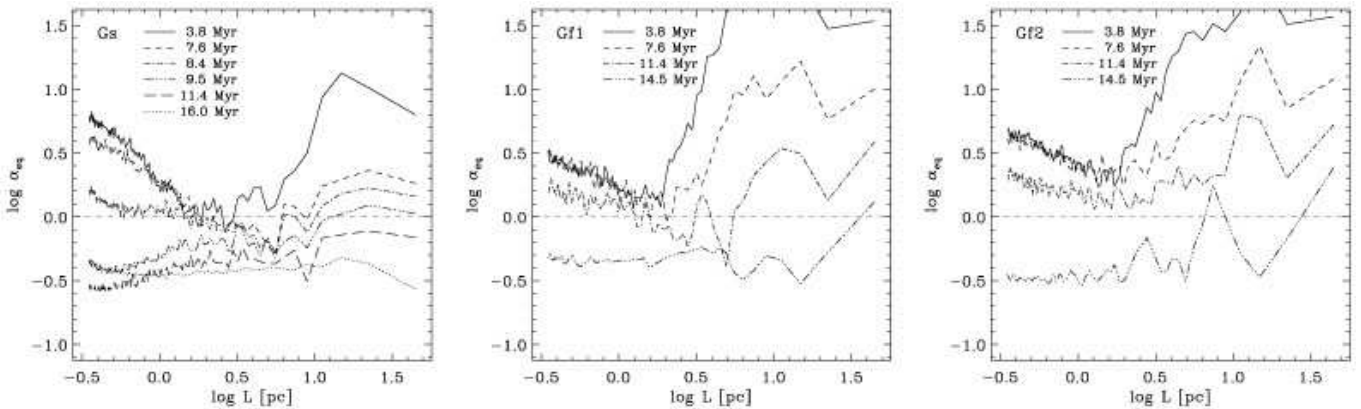}
  \caption{\label{f:virspec}The equipartition parameter $\alpha_{eq}$ against scale, for the three gravity
           models Gs, Gf1 and Gf2 as indicated in the panels. The line styles stand for
           the times at which $\alpha_{eq}$ has been measured. All models have been cut at a scale
           of $0.3$~pc, although the numerical resolution is a factor of approximately 3 (6) higher
           for model Gs (Gf1, Gf2).}
\end{figure*}

Figure~\ref{f:virpar} allows only an indirect statement about the scale-wise evolution
of $\alpha_{eq}$. A more accurate measure is the 
ratio of the respective Fourier spectra of 
the kinetic and internal energy over the gravitational energy. This yields $\alpha_{eq}(L)$, 
a scale-dependent measure of the cloud's stability against gravity (Fig.~\ref{f:virspec}). 
As in Figure~\ref{f:virpar},
we show the three gravitational models, but now at various times. Note that we do not select for
gas in the cloud or gas above a density threshold. This is because masking leads to additional
structure, which in turn results in ``noise'' signals in the Fourier spectra. To facilitate
an easier comparison to spatial scales we plot the scale-dependent $\alpha_{eq}$ against
physical length scale, rather than against wave number. Since $\alpha_{eq}$ is only a rough measure
of the system's energetics, the following discussion should be seen as a qualitative analysis.

At early times, the system is gravitationally stable on all scales. This is not surprising
since we are now looking pretty much at the whole box (we removed the $L=44$~pc mode, since this
would just be the mean). 
The small scales definitely collapse first: they are the first to fall 
beneath $\alpha_{eq}=1$  with increasing time. The minimum in $\alpha_{eq}$ around 
$L=1\dots 3$~pc at earlier times stems
from a ``conspiracy'' between the kinetic and potential energy scales: On larger scales 
($L\gtrsim 3$~pc), the kinetic energy of the large-scale inflow still dominates the energy budget
at early times, whereas on the small scales, the potential energy drops faster with decreasing
scale than the kinetic energy. The small structures forming due to thermal and dynamical fragmentation
did not have enough time yet to collect a significant amount of mass. 

In summary, we note that in all models local collapse wins over global collapse, i.e. the small scales
generated by thermal and dynamical fragmentation collapse first. Global collapse however occurs and
feeds more material into the already active ``star forming'' region. The clouds do not 
reach an 
``equilibrium stage'', but proceed from formation directly to local collapse and only then to global 
collapse. This result does not
depend on the presence or absence of stellar feedback, in contrast to the 
late-stage evolution of our model clouds, where feedback will have a deciding influence.

\subsubsection{Total and Core Mass Evolution}
\begin{figure}
  \includegraphics[width=\columnwidth]{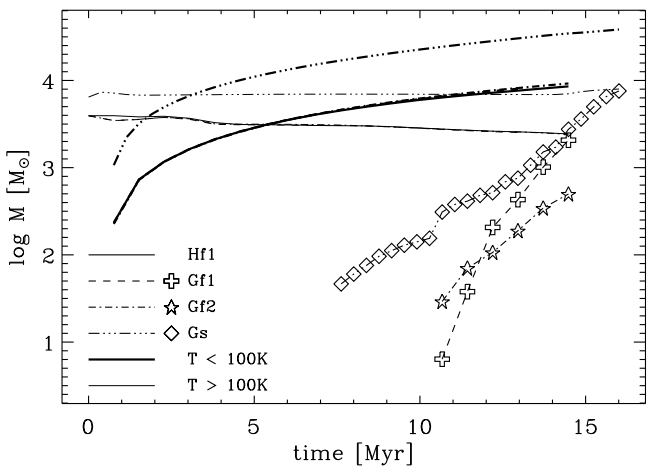}
  \caption{\label{f:allmasses}Mass history of all models. Thick lines denote 
           $M(T<100\mbox{K})$, thin lines stand for $M(T>100\mbox{K})$. Different
           line styles represent the four models. Symbols refer to the total mass
           in collapsing cores identified by CLUMPFIND (see \S\ref{ss:coreident}).}
\end{figure}

We have already seen the core mass history of model Gs (Fig.~\ref{f:masstime-Gs}).
Figure~\ref{f:allmasses} shows the total mass evolution of all models.
Thick lines refer to gas at $T<100$~K, which can be identified as cloud gas due
to the thermal instability. Thin lines denote the gas at $T>100$~K. 
The symbols stand for the total mass within collapsing cores for each model,
i.e. for $M_*$, the mass that constitutes the reservoir for star formation.

Note that the mass is scaled logarithmically, while the (cold) cloud mass evolves
linearly with time. Essentially, despite all the substructure, the colliding flows
form one cold slab (see also \citealp{2006ApJ...648.1052H}). Comparing models
Hf1, Gf1 and Gf2, gravity does not affect the global cloud mass: the inflows 
determine the mass in the cold gas phase (i.e. in the cloud), and thus the material available
for making stars. Model Gs differs from all others just because of the larger
extent of the collision site. Otherwise, its evolution regarding the cloud mass
is qualitatively similar.

The core masses evolve in parts nearly exponentially. This is a consequence not of
the individual mass accretion events, as Figure~\ref{f:masstime-Gs} demonstrates: those
are fairly linear with time. Rather, the non-linear evolution is a consequence of the
explosion of star formation activity once sufficient mass has  
accumulated. Comparing the mass in the cores, $M_*$, against the cloud mass $M_{cl}$ -- identified 
with $M(T<100\mbox{K})$ --, we note that over a large stretch of time, the ``star formation
efficiency'' $M_*/M_{cl} < 0.1$ -- the collapse is initially truly local. 

\subsubsection{Core Mass Distribution\label{sss:comafu}}
\begin{figure*}
  \includegraphics[width=\textwidth]{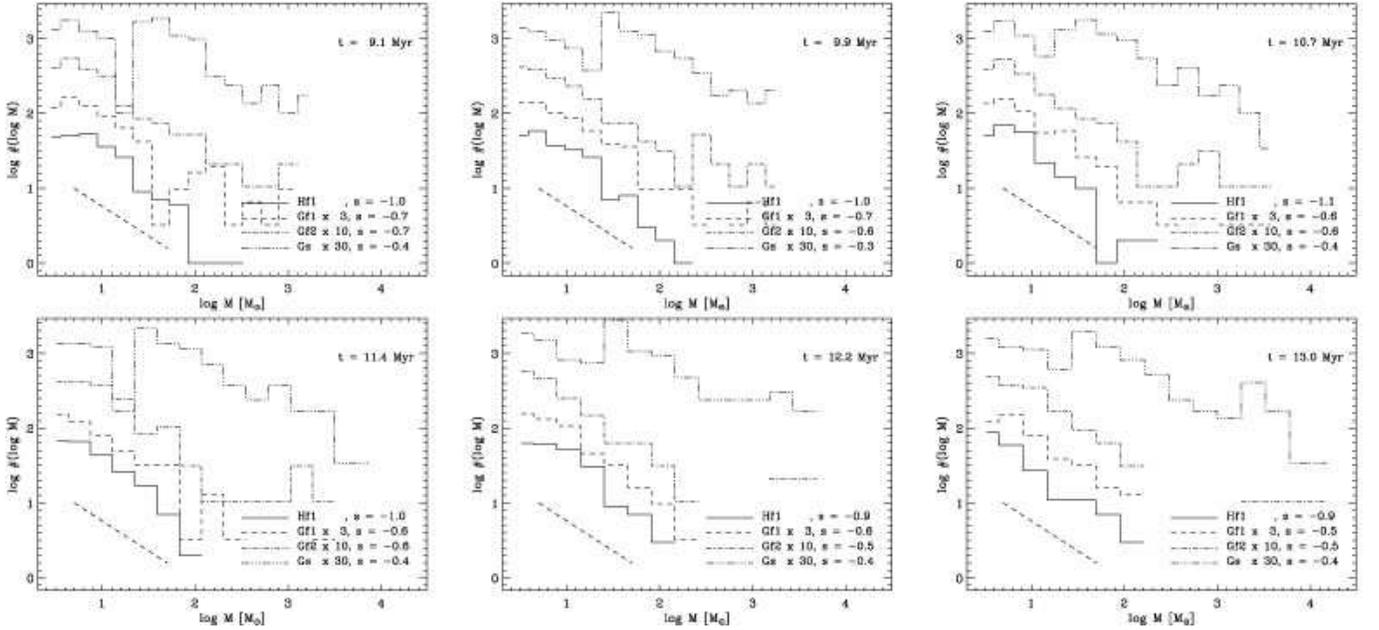}
  \caption{\label{f:coremassfunc}Time sequence of core mass distributions for all four
           models. Note that the distributions are shifted vertically by the factor
           indicated in the legend, to make them easier to identify.
           The Salpeter IMF would have a slope of $-1.35$. The dashed line denotes 
           $d(\ln N)/d(\ln M) = -0.8$.}
\end{figure*}

Since the thermal instabilities are the first fragmentation agent in the
colliding gas streams, will they set the core mass distribution in molecular
clouds? Figure~\ref{f:coremassfunc} shows the core mass distribution in logarithmic
mass intervals for all four models. The (fit) slopes $d\ln N/d(\ln M)= s$
are indicated in the legend. The Salpeter IMF would have an exponent of $s=-1.35$. 
While the slopes are approximately consistent
with observed core mass distributions
\citep{1996A&A...307..915K,2002A&A...384..225S}, 
we note that there is at most one decade in mass for the
fits. Moreover, these mass spectra are most likely affected by 
the stiffening of the equation of state for $n>10^5$~cm$^{-3}$ 
(\S\ref{ss:hydrocool}), i.e. low-mass objects may be under-represented.
This may cause a flattening of the spectra of the gravity-models
(Gf1, Gf2 and Gs) -- possibly visible in the lower right panel (13 Myr) of
Figure~\ref{f:coremassfunc} --, 
whereas model Hf1 shows a rather well-defined power law
down to the lowest masses at late times. 
Note that the cores contributing to Figure~\ref{f:coremassfunc} are not
necessarily self-gravitating. Over a period of 4~Myr, the spectra roughly keep
their shape, another indication that the fragmentation is mainly due to thermal
effects rather than gravity.

\subsection{Gas Dynamics}\label{ss:gasdyn}

\begin{figure}
  \includegraphics[width=\columnwidth]{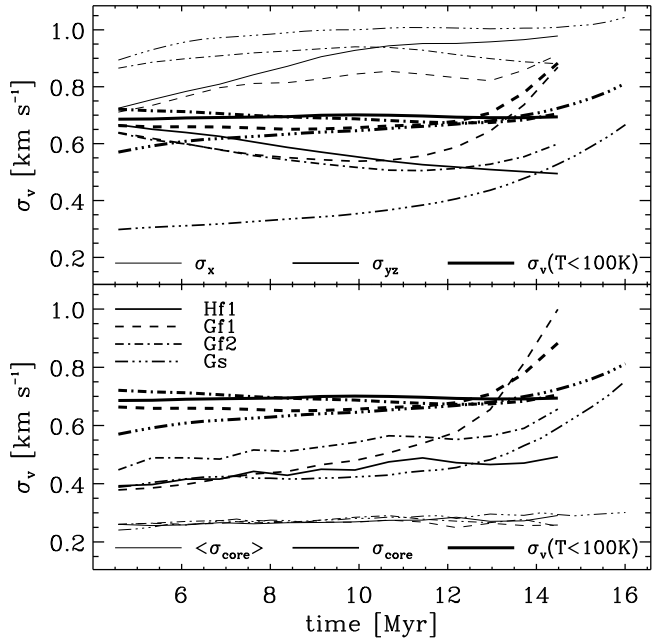}
  \caption{\label{f:veldisvt}{\em Top:} One-dimensional velocity dispersions against time,
           along the inflow direction ($\sigma_x$, thin lines), perpendicular to the 
           inflows ($\sigma_{yz}$medium lines), and total ($\sigma_v(T<100\mbox{K})$, thick
           lines), for all models.
           {\em Bottom:} One-dimensional velocity dispersions against time.
           We distinguish between the velocity dispersion within each core, 
           averaged over all cores ($\langle\sigma_{core}\rangle$, thin lines), the
           velocity dispersion of the gas taken over all cores ($\sigma_{core}$, medium lines),
           and the total velocity dispersion (as in top panel, thick lines).}
\end{figure}

Figure~\ref{f:veldisvt} summarizes the various line-of-sight velocity dispersions
(LOSVD) for the cold ($T<100$~K) gas, for all models. 
All LOSVDs are one-dimensional and density-weighted, e.g. the total LOSVD is 
\begin{equation}
  \sigma_v\equiv\left(\frac{\int\mbfv^2\,n\,dV}{3\int n dV}\right)^{1/2}.\label{e:losvd}
\end{equation}

The top panel shows the velocity dispersion along the inflow direction ($\sigma_x$),
the transverse velocity dispersion $\sigma_{yz}$, and the total dispersion.
For all models, $\sigma_x$ is highest, initially increasing slightly with time,
and later leveling off. The initial rise is due to the still acting NTSI, i.e.
the ripples in the interaction interface are still amplified, while at later
times, the NTSI is saturated, and global gravity takes over. The latter can 
be seen when comparing model Hf1 and Gf1, which are identical except that
Gf1 has self-gravity. For both, $\sigma_x$ initially evolves similarly, up
to $t\approx 7.5$~Myr. After that, $\sigma_x(\mbox{Gf1})$ starts to level off,
while $\sigma_x(\mbox{Hf1})$ still continues to rise: gravity constrains the
slab and suppresses the further growth of the NTSI. This is obvious in 
Figure~\ref{f:morph-Gf12p}. 

The transverse velocity dispersion $\sigma_{yz}$ (medium lines in top panel
of Fig.~\ref{f:veldisvt}) shows the influence of global gravity at later
times for models Gf1, Gf2 and Gs, while that of model Hf1 drops with time. 
For model Hf1, the total energy input of the inflows is balanced by radiative
losses and out-streaming material: the total velocity dispersion stays 
approximately constant with time. Not so for the self-gravitating models:
they all have increasing total velocity dispersions with time.

In the bottom panel of Figure~\ref{f:veldisvt}, the thick lines stand for
the total velocity dispersion again (as in the top panel). The thin lines
($\langle\sigma_{core}\rangle$) 
denote the velocity dispersion within each core, averaged over all cores.
This is the same quantity as the ``internal'' velocity dispersion as
discussed by \citet{2006ApJ...648.1052H}. Due to the strong radiative losses,
this velocity dispersion is subsonic (the sound speed in the cold gas
is $c_s\approx 0.3$~km~s$^{-1}$). In contrast, the velocity dispersion of 
all gas in the cores, $\sigma_{core}$ is slightly supersonic, and increases
with time due to global gravitational collapse. 

\subsection{Two Comments on Resolution}\label{ss:resolution}

\begin{figure*}
  \includegraphics[width=\textwidth]{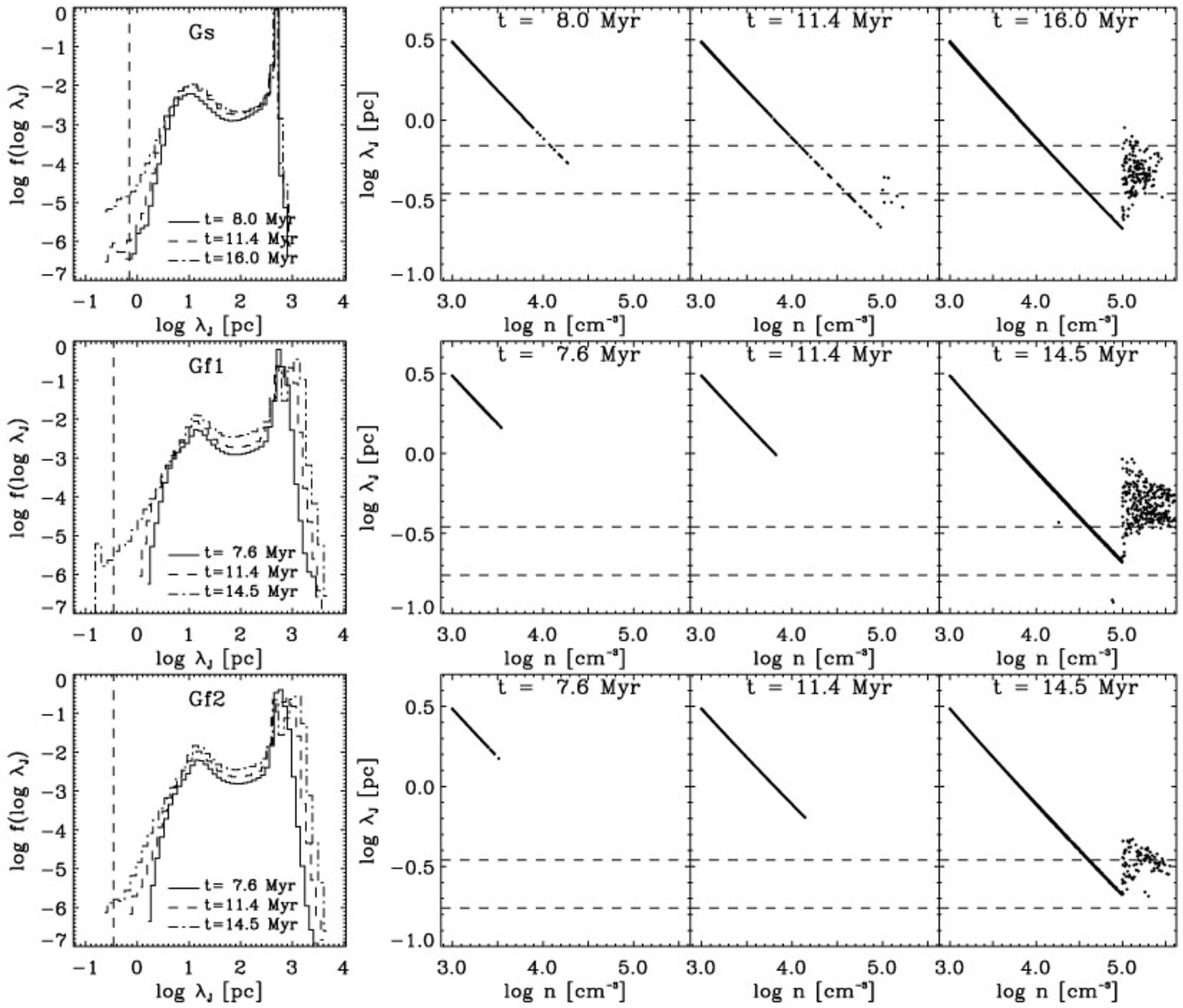}
  \caption{\label{f:jeanslength}Histograms of the Jeans length (left column)
          for all three self-gravitating models, and scatter plots of the Jeans
          length against density, for all models at three times. The dashed lines
          denote the \citet{1997ApJ...489L.179T} criterion with a safety factor of 2 and 4.}
\end{figure*}

The left column of Figure~\ref{f:jeanslength} 
shows the histograms of the Jeans length for all three
self-gravitating models. The dashed vertical line indicates the 
\citet{1997ApJ...489L.179T} criterion with a safety factor of 4,
mandating a minimum number of cells per Jeans length. We evaluate the criterion
locally, i.e. per cell. Clearly, with
increasing time, more and more cells fall below the resolution limit (to the
left of the vertical dashed line). A more detailed view is offered by the set of
panels on the right side of Figure~\ref{f:jeanslength}. They show the 
Jeans length in each cell against the corresponding density. Again,
dashed lines denote the Truelove limit, the upper line for 4 cells, the lower
one for 2. Although only a minor fraction of cells is unresolved, they
do exist. The strict correlation between $n$ and $\lambda_J$ for
$n<10^5$~cm$^{-3}$, and the scatter at larger densities, is a direct consequence
of the cooling curve: we are far up the isothermal branch, so that the
thermal timescales are much shorter than the dynamical timescales.
As discussed above, we switch off
the cooling for densities $n>10^5$~cm$^{-3}$, to prevent unphysically high
densities and subsequent numerical problems. The effective equation of state reverts
to adiabatic with $\gamma=5/3$ at that point, so that the temperature
increases due to the strong compressions.

There are several resolution criteria for thermally unstable systems (see e.g.
\citealp{2007A&A...465..431H} for a discussion).
The dynamically most stringent condition is to resolve the cooling
length $\lambda_c = c_s\,\tau_c$. If this length scale
is not resolved in the cold gas, the (isobaric) condensation mode of the thermal instability will be
underestimated (see \citealp{1965ApJ...142..531F} and \citealp{2000ApJ...537..270B}). 
The bulk of the cold gas resides at $40$~K for our cooling curve 
(see \citealp{2006ApJ...648.1052H}), and the cooling time scales in the cold gas are on the order
of $10^4$~years, so that the cooling length scale is $\lambda_c\approx 6\times 10^{-2}$~pc,
slightly beneath the nominal resolution for models Hf1, Gf1 and Gf2 (see \S\ref{ss:models}).
Underestimating the condensation mode of the thermal instability will result in fewer low-mass
fragments. In that sense, the core mass budgets (\S\ref{ss:colhist}) and the fragmentation
history in our models are conservative estimates: at higher resolution, more fragmentation 
-- and possibly earlier low-mass star formation -- is expected. 

%
%
\section{Discussion}\label{s:discussion}

\subsection{Global Collapse and Filament Formation}

Galactic star formation seems to prefer filamentary -- at least elongated --  
rather than spherical molecular clouds (\citealp{1979ApJS...41...87S};
\citealp{1997ApJ...474L.135C}; \citealp{2001ApJ...562..852H}; \citealp{2005A&A...440..151H};
\citealp{2007A&A...462L..17A}). 
The finite extent of the forming cloud opens two paths for filament formation. The first -- obvious -- one
is global collapse along the shorter axes. The second one -- less obvious -- arises from the fact that
the radial accelerations in a two-dimensional elliptical or circular cloud of uniform density diverge
at the edges, i.e. material at the edges experiences the strongest accelerations inwards, leading to 
a pile-up of gas at the edge, i.e to the formation of a filament \citep{2004ApJ...616..288B}.

Obviously, ``real'' clouds are three-dimensional. However, if clouds form in (laterally constrained)
colliding flows, they will have a finite extent, and more importantly, global gravity will not have
had sufficient time to lead to a centrally peaked density profile. The strong thermal instabilities lead
to an initially rather thin sheet, only broadened by the dynamical instabilities. Thus, to zeroth
order, in such a scenario the sheet-approximation is quite reasonable. 
The numerical models of \citet{2007ApJ...657..870V}
and the global collapse pattern of models Gf1 and Gf2 above support this interpretation. In fact, the 
dynamical instabilities will enhance a ``crumpling'' of the cloud in the lateral directions once
global gravity dominates. As the discussion in \S\ref{sss:global} shows however, stars will have formed
locally by then.

It is reasonable to assume that the (idealized) inflows implemented in our models will not have
a circular cross section. In galaxy mergers or in the collisions of super-shells in the LMC, 
the thickness of the disk would limit the thickness (vertical extent) of the flow. A similar
assumption seems valid for spiral shocks in the Galaxy, where the disk potential would again
lead to a flattening of the inflows (e.g. \citealp{2006MNRAS.371.1663D}).

\subsection{Formation of Massive Cores}
Ripples in the flow collision interface can focus the instreaming gas,
leading to a very efficient mechanism to form massive cores 
(Figs.~\ref{f:morph-Gs}, \ref{f:masstime-Gs}). 
$10$~Myr after flow collision, the first core
in model Gs has a mass of approximately $150$~M$_\odot$ and a diameter
of $1$~pc, corresponding to a column density of $3.6\times10^{22}$~cm$^{-2}$.
Likewise, after $10$~Myr, the mean column density in the box along the
inflow direction will be $N_{tot} = 3\mbox{ cm}^{-3}\times(44\mbox{ pc}+7.9\mbox{ km}
\mbox{ s}^{-1}\times 10\mbox{ Myr}) = 1.1\times 10^{21}$~cm$^{-2}$. The dynamical
focusing -- helped along by cooling and eventually gravity -- leads to an
excess of column-density by a factor of $\approx 30$. 

This is as good a place as any to remind the reader that we are leaving out the
crucial step of molecule formation in our models. The first appearance of 
molecules sets the clock for the lifetime of the cloud. We can
only mimic this by arguing that once column densities of $\approx 10^{21}$~cm$^{-2}$
are reached, we regard the cloud as ``molecular''. \citet{2001ApJ...562..852H} point out
that this column density is of the same order as the critical column density
for gravity to become dominant. 

With the NTSI as driving  mechanism, the cores will
form at a certain distance from the bulk of the cloud (Fig.~\ref{f:morph-Gs}). 
This distance is basically given by the amplitude the NTSI has reached up to 
that point. The somewhat peculiar location could have profound repercussions
on the effect of feedback from the massive stars forming in the 
core: the winds and the expanding HII regions or supernovae might
lead to a further compression of the already massive cloud next to the young
stars, thus triggering further star formation. In contrast, if the 
stars were located inside the bulk of the cloud, the stellar feedback
could be expected to provide partial support to the cloud, or more
likely disperse the lower-density regions.

\citet{2005ApJ...635.1062K} analyzed the molecular cloud population 
of the starburst galaxy M82, arguing that in many cases, star formation
seems to proceed from the outside inwards, i.e. that the regions of
massive star formation (indicated by HII regions and supernova remnants)
are to be found at the edges of the molecular clouds. They argue that
a sudden increase in external pressure (e.g. a shock wave traveling
through the cloud) triggers star formation in a previously existing
cloud. The dynamical focusing effect discussed here might offer
another explanation, which then would imply that the central molecular
cloud is still forming -- an alternative which may be even more attractive in view
of the notorious problem of stabilizing an object of many Jeans masses
against gravitational collapse (\citealp{2004ApJ...616..288B}; \citealp{2007RMxAA..43..123B}). 

\begin{figure}
  \begin{center}
  \includegraphics[width=\columnwidth]{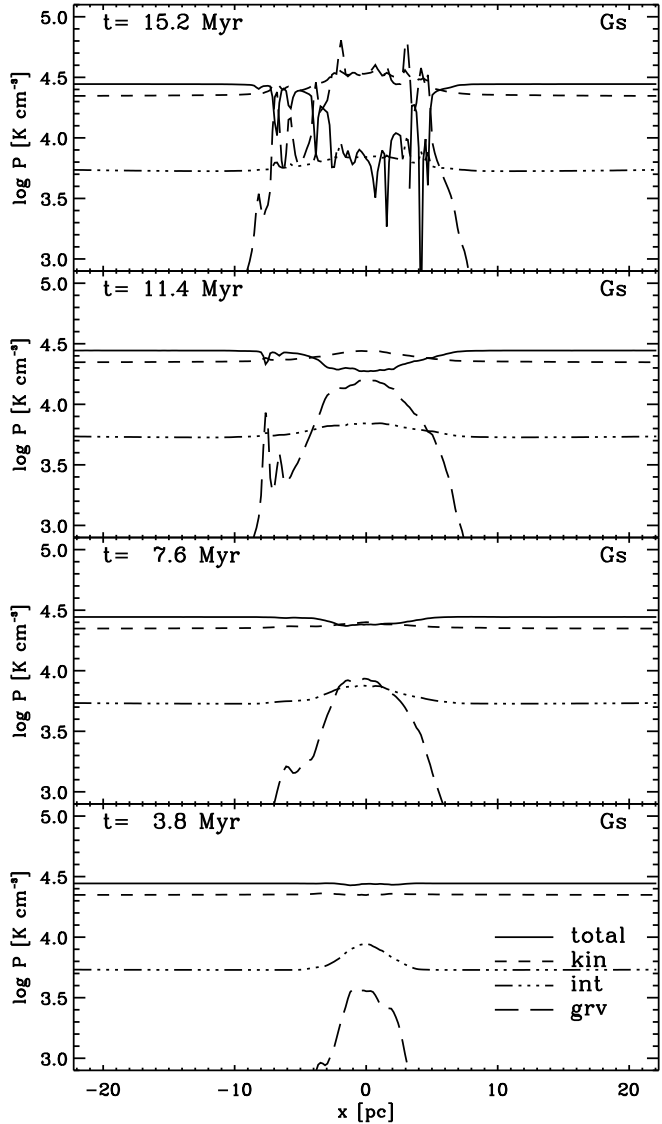}
  \end{center}
  \caption{\label{f:pressprof-Gs}Laterally averaged pressure (energy density) 
          profiles for 
          model Gs, at times as indicated in panels. At $11$~Myr, gravity
          has a noticeable global effect. Note that the internal pressure
          (dash-3-dot) varies only mildly.}
\end{figure}

The role of the ``external pressure increase'' in our models is taken over
by the dynamical focusing, i.e. by the excess ram pressure at the troughs
of the perturbed sheet. As Figure~\ref{f:pressprof-Gs} demonstrates, the
{\em thermal} pressure does not vary strongly. 

Thus, the gas still
evolves sub-isothermally, i.e. the effective adiabatic exponent
$\gamma < 1$. 

At late times (top panel, $15.2$~Myr), kinetic and gravitational pressure
are in equipartition. Once again, we emphasize that this does not mean that the
system is in equilibrium: the kinetic energy just follows the gravitational energy
\citep{2006MNRAS.372..443B}.

\subsection{Early Fragmentation and Core Mass Distributions}
The core mass distributions (Fig.~\ref{f:coremassfunc}) are consistent 
with observations \citep{1996A&A...307..915K,2002A&A...384..225S}, although they are flatter than
the Salpeter IMF. \citet{2007A&A...462L..17A} quote a core mass distribution
for the Pipe nebula close to the Salpeter IMF, however, as they point out, the cores
that they are using for analysis are probably the direct progenitors of stars -- 
a stage we cannot hope to reach with the models presented here.
Bearing in mind that the distributions 
for models Gf1, Gf2 and Gs are probably slightly too flat for numerical reasons
(\S\ref{sss:comafu}), their similarity to the non-gravitating model Hf1
substantiates the claim that the density substructure -- specifically the
core mass distribution -- in molecular clouds could very well arise 
very early on during their formation (\citealp{2007A&A...462L..17A};
\citealp{2007A&A...465..445H}) -- driven more by thermal than by gravitational fragmentation.

However, we caution that we have not followed the cloud evolution through
the stage of molecular hydrogen formation, an investigation we defer to a later paper.
Finally, we again emphasize the difficulty of determining mass spectra
from simulations with limited dynamic range. 

\subsection{Star Formation Duration and Stellar Age Spread}

From Figure~\ref{f:masstime-Gs} one would infer an 
``age spread'' of the stellar population in the complex of
up to $8$~Myr. This of course assumes that massive, self-gravitating
cores will sit around for several Myr without forming massive stars
whose energy input will tend to disrupt the cloud - a highly improbable
supposition.  While this aspect of our simulation is quantitatively
unrealistic, it does suggest the following qualitative points:

(1) Even in a completely dynamic simulation without magnetic or driven
turbulent support, the overall timescales of star formation can be
longer than the local collapse times.  Putting this another way,
the age spread in a star-forming region is an {\em upper limit}
to the timescale of local collapse; it is only equal to the local
collapse time if all star formation is globally synchronized, which
is less and less plausible on larger and larger scales.

(2) The large-scale flow picture of star-forming cloud accumulation is
attractive in that it allows for the formation of stars over timescales (1-2 Myr)
short compared with lateral crossing times (10-20 Myr), as observed \citep{2001ApJ...562..852H}.
This does not mean that a region of space may not have a significant
age spread.  In the paradigm we are pursuing, where non-linear fluctuations 
are important, it is plausible that a few large perturbations might collapse
long before more general star formation ensues, as we see in our simulations.
Although we form mostly massive cores, other initial conditions -- and higher
numerical resolution (see \S\ref{ss:resolution}) -- may lead
to the formation of a few low-mass stars initially from the low-probability
high density tail of perturbations, with the smaller, more frequent perturbations
constituting the bulk of the star formation at a later time.  Again, such
age spreads provide no constraint on local collapse timescales.

\subsection{The Role of Magnetic Fields}

Magnetic fields are not included in the models presented here.
Based on two-dimensional simulations by 
\citet{1995ApJ...441..702V}, \citet{2001ApJ...562..852H} envisaged 
the large-scale converging flows preferentially running along 
(dynamically dominant, e.g. \citealp{2005ApJ...624..773H}) magnetic
field lines in the diffuse ISM, arguing that if the field lines 
were oriented perpendicularly, the stiffening of the equation of 
state due to the increased magnetic pressure would counter the 
lowering of the effective adiabatic index due to (atomic line) 
cooling (see also \citealp{2004ApJ...612..921B}). Although
there are problems with this scenario (for one, substructures
can form even in the case of perpendicular field lines 
and lead to further fragmentation [e.g. 
\citealp{2007ApJ...665..445H}], and then, this scenario is
clearly motivated by two-dimensional simulations, not allowing
for interchange modes in the instabilities. These usually grow
at least at the hydrodynamical rate, see e.g. \citealp{2007arXiv0707.1022S}),
we will adopt it for the current study. Its consequence is that since
material is being piled up along the field lines, the mass-to-flux ratio
would easily exceed the critical value for magnetical dominance 
\citep{1976ApJ...210..326M} after a few Myr, thus rendering the magnetic
fields unimportant for the local gravitational collapse.
It is true that the global dynamics of the cloud especially during 
its early evolutionary stages could be fundamentally influenced by the
presence of magnetic fields. We will defer this important problem to 
a future paper.

%
%

\section{Summary}\label{s:answers}

In an extension of our previous work \citep{2005ApJ...633L.113H,2006ApJ...648.1052H}
we presented three-dimensional models of molecular cloud formation in colliding flows
including self-gravity. At an effective resolution of $512^3$ they currently are the 
most highly-resolved models of this kind. We used a fixed-grid code and a non-periodic 
Poisson-solver. The initial conditions emphasized the effects of global versus local 
gravity on the newly forming cloud.

The model resolution did not allow us to follow the collapsed cores and to study their properties.
Stellar feedback was not included. Thus, as time increases, the global evolution of
the molecular cloud gets less and less realistic. Furthermore, we cannot
make statements about the lifetime of the molecular cloud.
Bearing these caveats in mind, we found that:

(1) Any perturbation in the colliding flows is strongly amplified by a combination of
thermal and dynamical instabilities. Specifically, the thermal effects lead to local
high-density fragments that subsequently collapse before the (still forming) cloud
can collapse globally. Thus, cloud formation in colliding flows allows the rapid onset of
local star formation while evading the problem of globally collapsing clouds.
This is consistent with earlier findings \citep{2004ApJ...616..288B}
that local collapse can only 
win over global collapse in the presence of early non-linear density perturbations
(Fig.~\ref{f:virspec}).

(2) Even in the highly dynamical environment of the colliding flows, an 
elongated finite cloud under global gravity can form one or more filaments by
collapsing along its shorter axes. This lateral collapse 
opens up a further mass reservoir for star formation 
(Figs~\ref{f:morph-Gf12l}, \ref{f:morph-Gf12p}).

(3) Dynamical focusing, i.e. the deflection of incoming gas due to ripples in the 
interface between colliding gas streams, leads to very efficient high-mass core formation
(Figs~\ref{f:morph-Gs}, \ref{f:masstime-Gs}): core masses of a few $100$~M$_\odot$ are
reached within $\approx 10$~Myr.

(4) The clouds are not in any state of equilibrium at any time. Different scales submit
to gravitational collapse at different times, with the small scales going first
(Figs~\ref{f:virpar}, \ref{f:virspec}). Still, the global equipartition parameter 
(eq.~\ref{e:alphaeq}) ranges around $1$ within a 
factor of a few -- consistent with
observations when taking into account the observational uncertainties.

(5) The similarity of the core mass distribution for models with and without 
self-gravity and their robustness with time
indicate that it might be set very early on during
cloud formation. Thus, the core mass distribution would be a consequence of 
fragmentation due to (magneto-)hydrodynamical instabilities and cooling rather 
than due to self-gravity (Fig.~\ref{f:coremassfunc}).

(6) The clouds become globally unstable in the absence of feedback, leading to an exploding
``star formation efficiency'' (in quotes, since we only can form cores in our models, 
but not stars). The turbulence imparted by the (continuing inflows) does not suffice to
balance global gravity. As mentioned in \S\ref{s:introduction} and above, the 
absence of feedback in our models does
not allow us to test how long the clouds will survive. However, our models show that
the clouds will not disperse ``on their own accord'' in the background flows, i.e without
the help of stellar feedback.


\acknowledgements
Computations were performed at the NCSA (AST040026, AST060031) and on the 
local resources at U of M including the 64-processor cluster Star,
perfectly administered and maintained by J.~Hallum.
This work was supported by NASA grant NNG06GJ32G and the University of Michigan.
It has made use of the NASA Astrophysics Data System.

%
%

\bibliographystyle{apj}
\bibliography{./references}

\end{document}